# A uniform metal distribution in the intergalactic medium of the Perseus cluster of galaxies

Norbert Werner[1,2], Ondrej Urban[1,2,3], Aurora Simionescu[1,2,4], Steven W. Allen[1,2,3]

[1]Kavli Institute for Particle Astrophysics and Cosmology, Stanford University, 452 Lomita Mall, Stanford, CA 94305-4085, USA
[2]Department of Physics, Stanford University, 382 Via Pueblo Mall, Stanford, CA 94305-4060, USA
[3]SLAC National Accelerator Laboratory, 2575 Sand Hill Road, Menlo Park, CA 94025, USA
[4]Institute of Space and Astronautical Science (ISAS), JAXA, 3-1-1 Yoshinodai, Chuo-ku, Sagamihara, Kanagawa 252-5210, Japan

**Most of the metals (elements heavier than helium) ever produced by stars in the member galaxies of galaxy clusters currently reside within the hot, X-ray emitting intra-cluster gas. Observations of X-ray line emission from this intergalactic medium have suggested a relatively small cluster-to-cluster scatter outside of the cluster centers[1,2] and enrichment with iron out to large radii[3,4,5], leading to the idea that the metal enrichment occurred early in the history of the Universe[3]. Models with early enrichment predict a uniform metal distribution at large radii in clusters, while late-time enrichment, favored by some previous studies[6,7], is expected to introduce significant spatial variations of the metallicity. To discriminate clearly between these competing models, it is essential to test for potential inhomogeneities by measuring the abundances out to large radii along multiple directions in clusters, which has not hitherto been done. Here we report a remarkably uniform measured iron abundance, as a function of radius and azimuth, that is statistically consistent with a constant value of $Z_{Fe} = 0.306 \pm 0.012$ Solar out to the edge of the nearby Perseus Cluster. This homogeneous distribution requires that most of the metal enrichment of the intergalactic medium occurred before the cluster formed, likely over 10 billion years ago, during the period of maximal star formation and black hole activity.**

Between 2009 and 2011, we obtained a total of 84 observations of the Perseus Cluster with the Suzaku X-ray satellite, as a Key Project for that mission. The pointings covered eight azimuthal directions from the cluster center out to an offset angle of 2º, with a total exposure time of over 1 million seconds. We have analyzed the data from all three functioning X-ray Imaging Spectrometers, extracting spectra from annuli centered on the cluster center. We modeled the spectra from each of the 76 independent regions as a single-temperature thermal plasma in collisional ionization equilibrium, with the temperature, iron abundance, and spectrum normalization included as free parameters[8].

The measured radial and azimuthal variation of the iron abundance of the hot intra-cluster medium (ICM) out to the edge of the Perseus Cluster along the eight different directions is presented in Fig. 1. We define this `edge' as $r_{200}$, the radius within which the mean enclosed mass density of the cluster is 200 times the critical density of the Universe at the cluster redshift (for the Perseus Cluster $r_{200} = 1.8$ Mpc, corresponding to 82 arcmin[4]). We have tested the statistical uniformity of the measured iron abundance at radii $r > 400$ kpc (i.e. beyond the central metallicity peak associated with the brightest cluster galaxy[4,9]) by modeling the data from all radii and azimuths with a constant abundance. The measured chi-square value of 65.8 for 75 degrees of freedom is consistent with the null hypothesis of a constant iron abundance with a value of $Z_{Fe}$ = 0.306±0.012 Solar (68% confidence limit; we adopt a Solar iron abundance of $3.24\times10^{-5}$ relative to hydrogen by number[10]). This abundance is also consistent with the mean value measured at intermediate radii (~$0.2$-$0.5r_{200}$) for a sample of 48 clusters previously observed with XMM-Newton[2]. The iron abundance values measured at large radii in the less massive Virgo Cluster are lower, although these are likely biased by the multi-temperature structure of the gas in that system[5]. The azimuthally averaged iron abundance profile for the Perseus Cluster is shown in Fig. 2.

Enrichment scenarios in which the ICM is predominantly enriched with metals after the formation of clusters predict a non-uniform metal distribution, which is inconsistent with the observations presented here. Ram-pressure stripping[11] would result in a negative gradient in the metal abundance profiles out to large radii, as well as significant azimuthal variations, with higher metal abundances expected along directions connecting to surrounding large scale structure filaments[12]. Furthermore, metals ejected by galaxies at later times should roughly follow the distribution of galaxies, producing an approximately constant metal mass to light ratio, which is also inconsistent with observations[13]. Although large scale sloshing motions[14] may be able to mix metals to some degree, the strong gradients in the entropy distributions of cluster atmospheres[15], including Perseus[4,8], make the ICM convectively stable, prohibiting the efficient mixing of metals that are initially non-uniformly distributed across large radial ranges.

The observed uniform iron abundance distribution at large radii in the Perseus Cluster requires early enrichment of the intergalactic gas in the proto-cluster environment. This enrichment was most likely driven primarily by galactic winds[16] - energetic outflows of metal enriched gas - which are expected to be strongest at epochs around the peak of star formation[17] and active galactic nuclei (AGN) activity[18] (redshifts $z \sim$ 2-3, or lookback times of 10-12 billion years). The combined energy of supernova explosions[16] and AGN[19] must have been strong enough to expel most of the metals from the galaxies at early times, and enrich and mix the intergalactic gas. This gas was later accreted by clusters and virialized (increasing its entropy through shock heating) to form the present ICM.

The total iron mass in the ICM of the Perseus Cluster, calculated from the measured iron abundance and ICM density profiles[8], is about 50 billion Solar masses, with about 60% residing beyond $0.5r_{200}$. The dominant fraction of this iron was likely supplied by type Ia supernovae (SNIa), which are thought to have produced 60-90% of iron in the Perseus Cluster[13] depending on the assumed supernova yields. Based on our measurements, we estimate that at least 40 billion SNIa contributed to the chemical enrichment of the proto-cluster environment that later formed the Perseus Cluster. For such a large iron mass to be expelled by the winds from the galaxies at early times, a significant fraction of these SNIa must have exploded shortly after the epoch of peak star-formation. This is consistent with recent findings based on SNIa delay time distributions[20,21], which imply that a large fraction of SNIa explode less than ~$5\times10^8$ years after the formation of the progenitor binary system (prompt SNIa). SNIa with longer delay times will continue contributing to the enrichment of the ICM after clusters virialize, and are expected to be partly responsible for the metallicity peaks surrounding the brightest cluster galaxies in the centers of many clusters[22].

A unique prediction of the early enrichment scenario is that essentially all galaxy clusters with masses comparable to the Perseus Cluster should have homogeneous iron abundance distributions of about one-third Solar at large radii. Also, contrary to some initial findings[6,7], there should be no substantial redshift evolution in the ICM metallicity outside of the central regions of clusters, out to $z \sim 2$. The presence and strength of such evolution are a matter of ongoing debate[2,23,24]. The observed large iron abundance of the high-entropy gas in the outskirts of the Perseus Cluster is also consistent with the idea that the highest-energy cosmic rays (above a few $10^{18}$ eV) may be primarily iron nuclei[25] accelerated by cluster formation shocks[26]. Additionally, because much of the metal rich ICM seen at large radii has been accreted from the surrounding large scale structure filaments, our model predicts that the tenuous warm-hot intergalactic medium (WHIM) permeating the cosmic web, in which up to half of the baryons in the Universe currently reside[27,28], is likely to be substantially enriched in metals. The same is true for the hot circumgalactic gas accreted by galaxies from the intergalactic-medium.

**Acknowledgements** The authors are grateful for discussions with other members of the Perseus Cluster Suzaku Key project collaboration, as well as with Yu Lu, Paul Simeon, and Roger Blandford. This work was supported by the Suzaku grants NNX09AV64G, NNX10AR48G, NASA ADAP grant NNX12AE05G, and by the U.S. Department of Energy under contract number DE-AC02-76SF00515.

**Author Contributions** NW led the writing of the manuscript. OU reduced and analyzed the data. AS and NW contributed to the data analysis. SWA contributed to the writing of the manuscript and is the principal investigator of the Suzaku Key Project data. All authors discussed all results, developed the interpretation, and commented on the manuscript.

**Competing Interests** The authors declare that they have no competing financial interests.

**Correspondence** Correspondence and requests for materials should be addressed to NW (email: norbertw@stanford.edu).

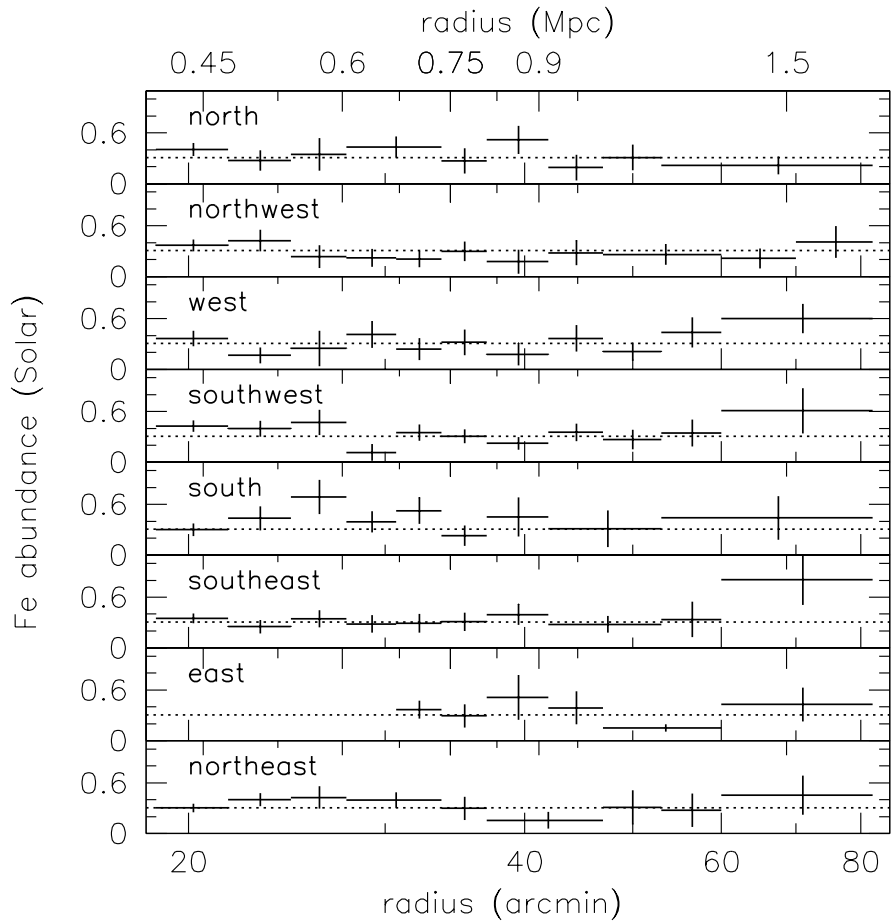

Figure 1: **Iron abundance profiles expressed in Solar units[10] measured along eight different directions in the Perseus Cluster.** At the temperature of the Perseus Cluster, the iron abundance measurement is driven by the Fe–K lines. Compared to the Fe-L lines, these lines are significantly less prone to systematic biases associated with the presence of multi-temperature gas. The spectral extraction regions span a relatively small range in radius and azimuth, where the plasma can be well approximated as single-temperature. There is no evidence for a bias in the temperature measurements due to gas clumping[8]. We have verified that spectral fits with the Fe-L line complex (0.8-1.5 keV energy range) excluded give consistent results to those in the full 0.7-7 keV band. Collisional ionization and electron-ion equilibrium are likely to hold out to $r_{200}$ in the Perseus Cluster, where the equilibration timescales are only ~$3.5 \times 10^8$ yr and ~$7.3 \times 10^8$ yr, respectively. Within a radius $r \sim 400$ kpc (~$0.2 r_{200}$; not shown here) metal enrichment is strongly

influenced by the brightest cluster galaxy and the metallicity is centrally peaked[4,9]. Along the eastern direction, large scale sloshing motions are present and are uplifting the ICM from smaller radii out to $r \sim 650$ kpc[14]. Thus we also do not present results for $r < 650$ kpc in this direction. Beyond these radii, the iron abundance measurements are consistent with a radially and azimuthally constant value of $Z_{Fe} = 0.306 \pm 0.012$ Solar, indicated by the dotted lines. Expressed in units of an older, and historically more commonly used Solar iron abundance value[29], our best fit constant iron abundance is $Z_{Fe} = 0.212 \pm 0.008$ Solar. The plotted error bars are 68% confidence intervals.

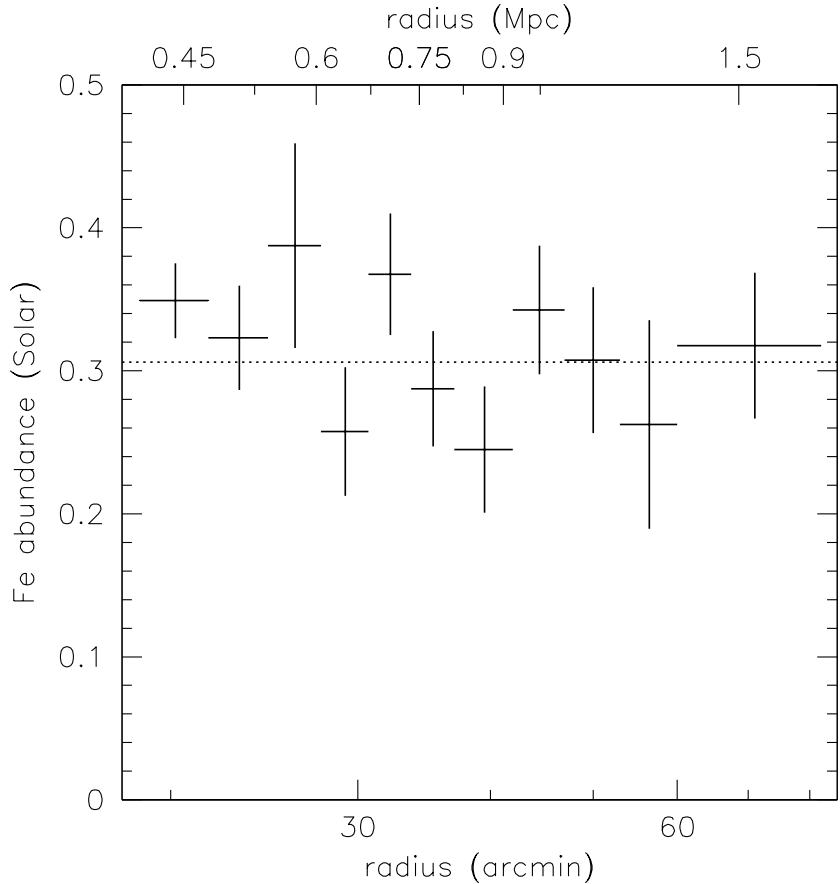

Figure 2: **Azimuthally averaged iron abundance profile for the Perseus Cluster**. The profile was determined by simultaneous fitting of the spectra at a given radius for all eight azimuthal directions. While the iron abundance was assumed to be the same in all directions in the fits, the temperatures and spectral normalizations were allowed to vary independently along the different azimuths. The dotted line shows the best fit constant iron abundance value of $Z_{Fe} = 0.306$ Solar. The plotted error bars are 68% confidence intervals. Our iron abundance values are lower than the value measured in the 650-1100 kpc range with XMM-Newton[13]; however, the XMM-Newton measurement suffers from significant systematic uncertainties due to the high and variable particle background of that instrument compared to Suzaku.